\documentclass[letter]{ptptex}

\usepackage{graphicx}
\usepackage{wrapft}
\newcommand{\sol}{M_{\odot}}
\newcommand{\rscript}[1]{\mbox{\scriptsize #1}}
\newcommand{\dOmega}{\mbox{d}\Omega}
\makeatletter
\@ifundefined{lesssim}{\def\lesssim{\mathrel{\mathpalette\vereq<}}}{}
\@ifundefined{gtrsim}{\def\gtrsim{\mathrel{\mathpalette\vereq>}}}{}
\def\vereq#1#2{\lower3pt\vbox{\baselineskip1.5pt \lineskip1.5pt
\ialign{$\m@th#1\hfill##\hfil$\crcr#2\crcr\sim\crcr}}}
\makeatother

\title{%
Gauge-Invariant Gravitational Wave Extraction from Coalescing Binary
Neutron Stars}

\author{
Mari \textsc{Kawamura}$^{1}$
and
Ken-ichi \textsc{Oohara}$^{2}$
}

\inst{%
$^1$ Graduate School of Science and Technology, Niigata University,
  Niigata, 950--2181, Japan\\
$^2$ Department of Physics, Niigata University,
         Niigata 950--2181, Japan\\
}

\abst{%
  We report application of a method for extracting gravitational waves
  to three-dimensional numerical simulation on coalescing binary
  neutron stars. We found the extracted wave form includes the
  componets corresponding to the quadrupole part in the Newtonian
  potential of the background metric, if it is monitored at a position
  not far from the central stars. We present how to eliminate it.}

\begin{document}

\maketitle

We are constructing computer codes on three-dimensional numerical
relativity.\cite{ONS97,ON99} At first we  used the conformal slicing
condition, in which the metric becomes the Schwarzschild one in the
outer vacuum region so that if the three-metric is split into the
Schwarzschild background and the perturbed parts, the latter can be
considered as the gravitational waves at the wave zone.\cite{SN92}
However, it was found that this slicing involves unstable modes and
long-term evolution of coalescing binary neutron star cannot be
followed.\cite{ONS97,SN95}
Then we started to construct a new code using the maximal slicing
condition. In this slicing, the perturbed part of the three-metric
includes gauge dependent modes and therefore we need gauge-invariant
wave extraction. Recently  gauge-invariant wave extraction methods
have been given as nonspherical perturbations of Schwarzschild
geometry.\cite{AE88,AE90,ADHS92,ABR97} In this letter, we report
application of a method based on them to three-dimensional general
relativistic simulation on coalescing binary neutron stars.

We use (3+1)-formalism of the Einstein equation and write the line
element as $ds^2 = - \alpha^2 dt^2 + \gamma_{ij} ( dx^i + \beta^i dt )
( dx^j + \beta^j dt)$. Outside of the star, we split the total
spacetime metric $g_{\mu\nu}$ into a Schwarzschild background and
non-spherical perturbation parts: $g_{\mu\nu} =
g^{(B)}_{\mu\nu}+h_{\mu\nu}^{(e)} + h_{\mu\nu}^{(o)}$, where
$g^{(B)}_{\mu\nu}$ is a spherically symmetric metric given by
\begin{equation}
  g^{(B)}_{\mu\nu} d x^{\mu} d x^{\nu} = 
  -N^2 d t^2 + A^2 d r^2 + R^2 (d \theta^2 + \sin^2 \theta d \phi^2)
\end{equation}
and $h^{(e)}_{\mu\nu}$ and $h^{(o)}_{\mu\nu}$ are even-parity and
odd-parity metric perturbations, respectively;
\begin{equation}
  h^{(e)}_{\mu\nu} = \sum_{lm}
  \left(\begin{array}{@{}c@{\ }c@{\
        }cc@{}} N^2H_{0lm}Y_{lm} & H_{1lm}Y_{lm} &
      h^{(e)}_{0lm}Y_{lm,\theta} & h^{(e)}_{0lm}Y_{lm,\phi} \\
      \mbox{sym} & A^2H_{2lm}Y_{lm} & h^{(e)}_{1lm}Y_{lm,\theta} &
      h^{(e)}_{1lm}Y_{lm,\phi} \\
      \mbox{sym} & \mbox{sym} &
      r^2 ( K_{lm}Y_{lm}+G_{lm} Y_{lm,\theta\theta}) &
      r^2G_{lm}X_{lm} \\
      \mbox{sym} & \mbox{sym} & \mbox{sym} &
      h^{(e)}_{33}
    \end{array}
  \right) ,
  \label{eq:heven}
\end{equation}
\begin{equation}
  h^{(e)}_{33} = r^2\sin^2\theta \left[ K_{lm}Y_{lm} + G_{lm} \left(
   Y_{lm,\theta\theta} - W_{lm} \right) \right],
\end{equation}
and
\begin{equation}
  h^{(o)}_{\mu\nu} = \sum_{lm}\left(
    \begin{array}{cccc}
      0 & 0 & -h^{(o)}_{0lm}Y_{lm,\phi}/\sin\theta &
      h^{(o)}_{0lm}Y_{lm,\theta} \sin{\theta} \\
      0 & 0 &
      -h^{(o)}_{1lm} Y_{lm,\phi}/\sin\theta &
      h^{(o)}_{1lm}Y_{lm,\theta}\sin{\theta} \\
      \mbox{sym} &
      \mbox{sym} & h^{(o)}_{2lm} X_{lm}/\sin\theta &
      -h^{(o)}_{2lm}W_{lm}\sin{\theta} \\
      \mbox{sym} & \mbox{sym} & \mbox{sym} &
      -h^{(o)}_{2lm}X_{lm} \sin{\theta}
    \end{array}
  \right),
\label{eq:hodd}
\end{equation}
where the symbol `sym' indicates the symmetric components,
$H_{1lm}$, $h^{(e)}_{0lm}$, $h^{(e)}_{1lm}$, $K_{lm}$, $G_{lm}$,
$h^{(o)}_{0lm}$, $h^{(o)}_{1lm}$, and $h^{(o)}_{2lm}$ are the
functions of $t$ and $r$; $Y_{lm}$ is the spherical harmonics,
$X_{lm}$ and $W_{lm}$ are given by
\begin{equation}
  X_{lm} = 2( Y_{lm,\theta\phi} - Y_{lm,\phi} \cot\theta) ,
  \quad
  W_{lm} = Y_{lm,\theta\theta} - Y_{lm,\theta} \cot\theta 
  - Y_{lm,\phi\phi}/\sin^2\theta .
\end{equation}
From the linearized theory about perturbations of the Schwarzschild
spacetime, the gauge invariant quantities $\Psi^{(o)}$ and
$\Psi^{(e)}$ are given by\cite{MON74}
\begin{equation}
  \Psi^{(o)}_{lm}(t,r) = \sqrt{2\Lambda(\Lambda-2)} N^2
  \left( h^{(o)}_{1lm} + r^2 ( h^{(o)}_{2lm}/r^2)_{,r} \right)/r
\end{equation}
and
\begin{equation}
  \Psi^{(e)}_{lm}(t,r) = -\sqrt{2(\Lambda-2)/\Lambda} \
  (4rN^2k_{2lm} + \Lambda rk_{1lm})/(\Lambda + 1 - 3N^2)
\end{equation}
for the odd and even parity modes, respectively, where
$\Lambda = l(l+1)$,
\begin{equation}
  k_{1lm} = K_{lm} + N^2 rG_{lm,r} - 2 N^2 h^{(e)}_{1lm}/r
\end{equation}
and
\begin{equation}
  k_{2lm} = H_{2lm}/(2N^2) - ( rK_{lm}/N^2 )_{,r}/\sqrt{N^2}.
\end{equation}
The quantities $\Psi^{(o)}$ and $\Psi^{(e)}$ satisfy the Regge-Wheeler
and the Zerilli equations, respectively.\cite{ZER24} Two independent
polarizations of gravitational waves $h_{+}$ and $h_{\times}$ are
given by
\begin{equation}
  h_{+} - \mbox{i} h_{\times} = \frac{1}{\sqrt{2}r} \sum_{l,m}
  (\Psi^{(e)}_{lm}(t,r) + \Psi^{(o)}_{lm}(t,r))_{-2}Y_{lm},
\end{equation}
where
\begin{equation}
  {}_{-2}Y_{lm} = ( W_{lm} - \mbox{i} X_{lm}/\sin \theta ) /
  \sqrt{\Lambda(\Lambda-2)}. 
\end{equation}
In numerical calculations, the functions $N^2(t,r)$, $A^2(t,r)$ and
$R^2(t,r)$ of the background metric are calculated by performing the
following integration over a two-sphere of radius $r$:\cite{ADHS92}
\begin{equation}
  N^2 = -\frac{1}{4\pi} \int g_{tt} \, \dOmega , \quad
  A^2 = \frac{1}{4\pi} \int g_{rr} \, \dOmega , \quad
  R^2 = \frac{1}{8\pi} \int \left( g_{\theta\theta} +
    \frac{g_{\phi\phi}}{\sin^2\theta} \right) \, \dOmega,
\end{equation}
where $\dOmega = \sin \theta \mbox{d}\theta \mbox{d}\phi$.  The
components of metric perturbations are
\begin{equation}
  H_{2lm}(t,r) = \frac{1}{A^2} \int g_{rr} Y^{\ast}_{lm} \, \dOmega ,
\end{equation}
\begin{equation}
  G_{lm}(t,r) = \frac{1}{\Lambda(\Lambda-2)} \frac{1}{R^2} \int \left[
  \left( g_{\theta\theta} - \frac{g_{\phi\phi}}{\sin^2\theta} \right)
  W^{\ast}_{lm} + \frac{2g_{\theta\phi}}{\sin\theta} \,
  \frac{X^{\ast}_{lm}}{\sin\theta} \right] \, \dOmega,
\end{equation}
\begin{equation}
  K_{lm}(t,r) = \frac{1}{2}\Lambda G_{lm} +\frac{1}{2R^2} \int \left(
  g_{\theta\theta} +\frac{g_{\phi\phi}}{\sin^2\theta} \right)
  Y^{\ast}_{lm}) \, \dOmega ,
\end{equation}
\begin{equation}
  h^{(e)}_{1lm}(t,r) = \frac{1}{\Lambda} \int \left(
  g_{r\theta}Y^{\ast}_{lm,\theta} + \frac{g_{r\phi}}{\sin\theta} \,
  \frac{Y^{\ast}_{lm,\phi}}{\sin\theta} \right) \, \dOmega,
\end{equation}
\begin{equation}
  h^{(o)}_{1lm}(t,r) = -\frac{1}{\Lambda} \int \left( g_{r\theta}
  \frac{Y^{\ast}_{lm,\phi}}{\sin\theta} -\frac{g_{r\phi}}{\sin\theta}
  Y^{\ast}_{lm,\theta} \right) \, \dOmega 
\end{equation}
and
\begin{equation}
   h^{(o)}_{2lm}(t,r) = \frac{1}{2\Lambda(\Lambda-2)} \int \left[
   \left( g_{\theta\theta} - \frac{g_{\phi\phi}}{\sin^2\theta} \right)
   \frac{X^{\ast}_{lm}}{\sin\theta} - \frac{2g_{\theta\phi}}{\sin\phi}
   W^{\ast}_{lm} \right] \, \dOmega ,
\label{eq:ho2}
\end{equation}
where $\ast$ denotes the complex conjugate.

We need angular integrals over spheres for constant $r$, such as
\begin{equation}
  \label{eq:IntdOmega}
  F(r_0) = \int_{r = r_0} f(x,y,z) \,\dOmega = \int f(r_0,\theta,\phi)
  \,\dOmega.
\end{equation}
If numerical simulation is performed using Cartesian coordinate system,
we need interpolation to obtain the values of $f(r_0,\theta,\phi)$ from
$f(x,y,z)$ at the grid points. It is, however, not easy to fully
parallelize the procedure on a parallel computer with distributed
memory.  We therefore rewrite Eq.(\ref{eq:IntdOmega}) as a volume
integral, namely,
\begin{equation}
  \label{eq:IntdV}
  F(r_0) = \int f(x,y,z) \delta(r-r_0) \, \mbox{d}^3 x =
  \lim_{a\rightarrow 0} \frac{1}{\sqrt{\pi} a r_0^2} \int F(x,y,z)
  e^{-(r-r_0)^2/a^2} \, \mbox{d}^3 x ,
\end{equation}
where $r = \sqrt{x^2 + y^2 + z^2}$. Numerical integral with $a =
\Delta x/2$ gives a good value to Eq.(\ref{eq:IntdV}), where $\Delta
x$ is the separation between grid points.

We performed numerical simulation for a coalescing binary consisting
of two identical neutron stars of mass $1.5\sol$ and evaluated the
gravitational waves. The details of our code will be shown
elsewhere\cite{KON04} but it is essentially the same as
Refs.~\citen{ON99} and \citen{ON02}. The lapse function and the shift
vector are determined by the maximal slicing and the pseudo-minimal
distortion conditions, respectively. We used uniform $475\times
475\times 238$ Cartesian grid with $\Delta x = 1\sol$ assuming the
symmetry with respect to the equatorial plane. As for an equation of
state, we use the ${\gamma}=2$ polytropic equation of state. The
initial rotational velocity is given so that the circulation of the
system vanishes. The ADM mass of the system is $2.8\sol$.

\begin{figure}[tbp]
  \begin{minipage}[t]{.49\textwidth}
  \centerline{\includegraphics[width=.95\textwidth]{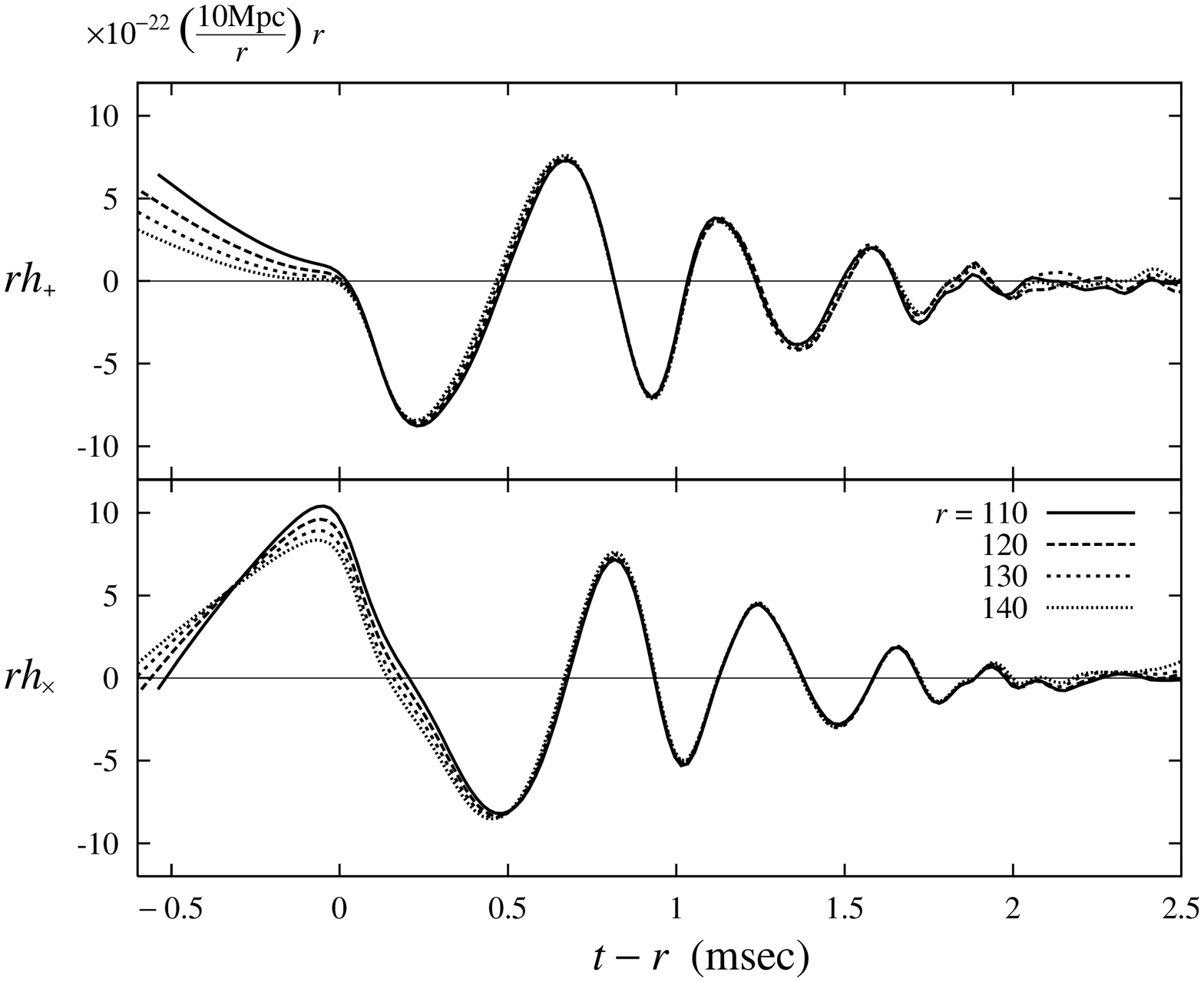}}
  \caption{Plots $r h_{+,\times}$ along $z$-axis at $r = 110 \sim
    140 \sol$ as a function of $t-r$.}
    \label{fig:wave0}
  \end{minipage}
  \hfill
  \begin{minipage}[t]{.49\textwidth}
  \centerline{\includegraphics[width=.95\textwidth]{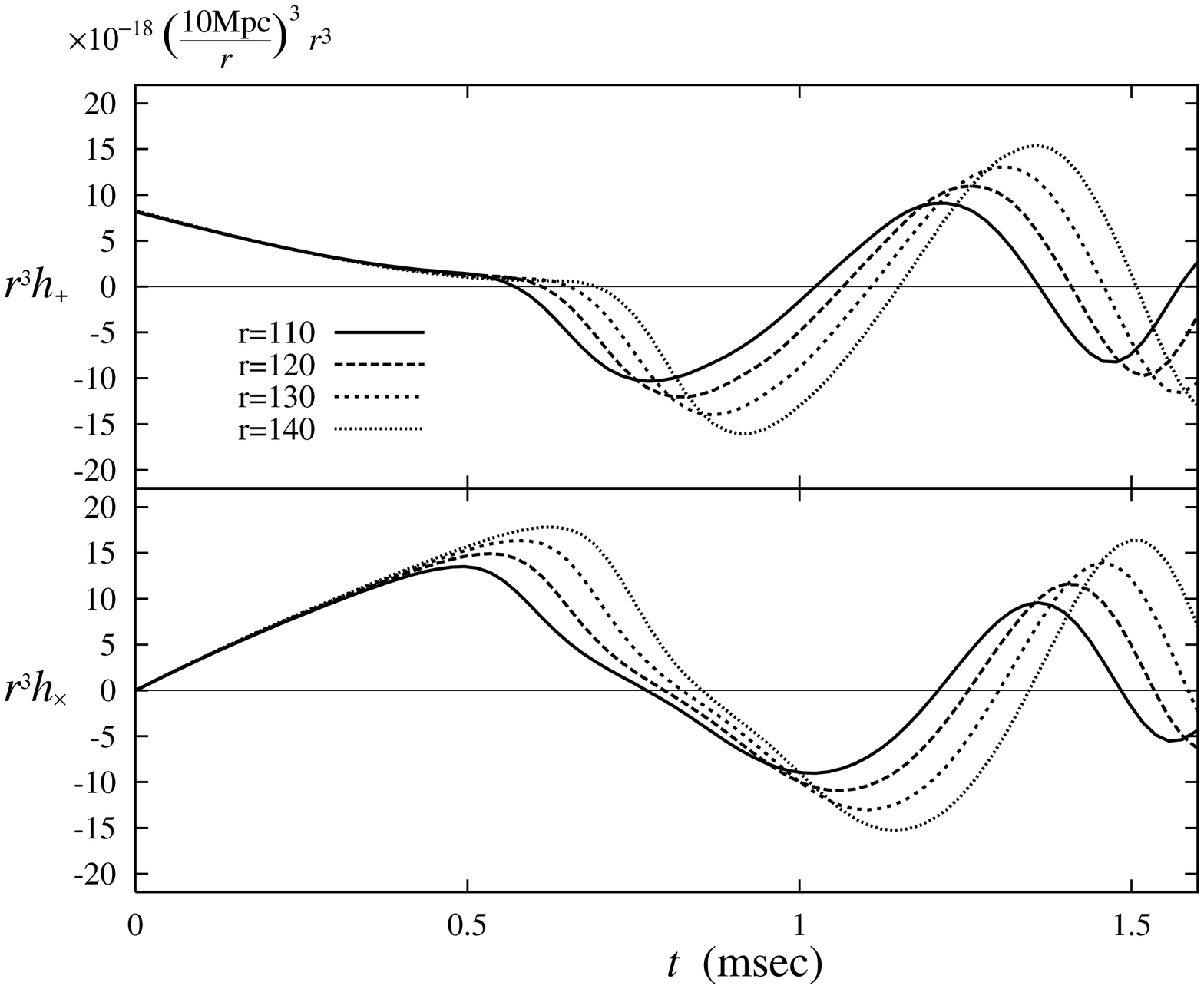}}
  \caption{Plots $r^3 h_{+,\times}$ as a function of $t$.}
    \label{fig:wave3}
  \end{minipage}
\end{figure}

Figure \ref{fig:wave0} shows the gravitational wave forms $rh_+$ and
$rh_\times$ on the $z$-axis evaluated at $r=110, 120, 130$ and $140
\sol$ as functions of the retarded time $t-r$.  The lines of
$rh_{+,\times}(t-r)$ estimated at $r = 110 \sim 140 \sol$ for $t-r
\gtrsim 0$ coincide with each other. Then the waves proportional to
$r^{-1}$ and propagating at the speed of light are extracted. For $t-r
\lesssim 0$, however, $h$ includes a non-wave mode proportional to
$r^{-3}$ as shown in Fig.~\ref{fig:wave3}, where $r^3 h$ is plotted as
a function of $t$. This mode corresponds to the quadrupole part in the
Newtonian potential of the background metric. It decreases fast as the
merger of stars proceeds. Since this mode is proportional to $r^{-3}$
while the wave mode is to $r^{-1}$, the former will be negligible if
the waves are monitored at a few times farther position. Now we can
eliminate the non-wave mode from $h_{+}$ and $h_{\times}$ using
Fourier transformation as follows:
\begin{itemize}
\item Assuming that total waves are expressed as a sum of wave parts
$F(t-r)/r$ and non-wave parts $G(t)/r^3$,
  \begin{equation}
    h(t,r)=\frac{F(t-r)}{r}+\frac{G(t)}{r^3}.
  \end{equation}
\item Fourier components of $h(t,r)$ are written as
  \begin{equation}
    h_{\omega}(r) = \frac{e^{-\rscript{i}{\omega}r}}{r}F_{\omega}(r) +
    \frac{1}{r^3}G_{\omega}(r).
  \end{equation}
  where
  \begin{equation}
    F_{\omega} \equiv \frac{1}{2\pi}
    \int F(t) e^{-\rscript{i}\omega t} \,\mbox{d}t
    \quad
    \mbox{and} \quad
    G_{\omega} \equiv \frac{1}{2\pi}
    \int G(t) e^{-\rscript{i}\omega t} \,\mbox{d}t.
  \end{equation}
\item From the values of $h_{\omega}(r)$ in different radial
  coordinates $r_1$ and $r_2$, $F_{\omega}$ can be given by
  \begin{equation}
    F_{\omega}=\frac{r_{2}^{3}h_{\omega}(r_2)-r_{1}^{3}h_{\omega}(r_1)}
    {r_{2}^{2} e^{-\rscript{i} \omega r_2}
      - r_{1}^{2} e^{-\rscript{i} \omega r_1}}.
  \end{equation}
\item By inverse Fourier transformation, we can get the gravitational
  waves, which do not include non-wave modes,
  \begin{equation}
    \label{eq:defhpx}
    h_{+}(t,r) - \mbox{i} h_{\times}(t,r) =\int
    \frac{e^{-\rscript{i} \omega r}}{r}F_{\omega}
    e^{\rscript{i} \omega t}\,\mbox{d}\omega.
  \end{equation}
\end{itemize}

\begin{figure}[tbp]
  \begin{minipage}{.49\textwidth}
  \centerline{\includegraphics[width=\textwidth]{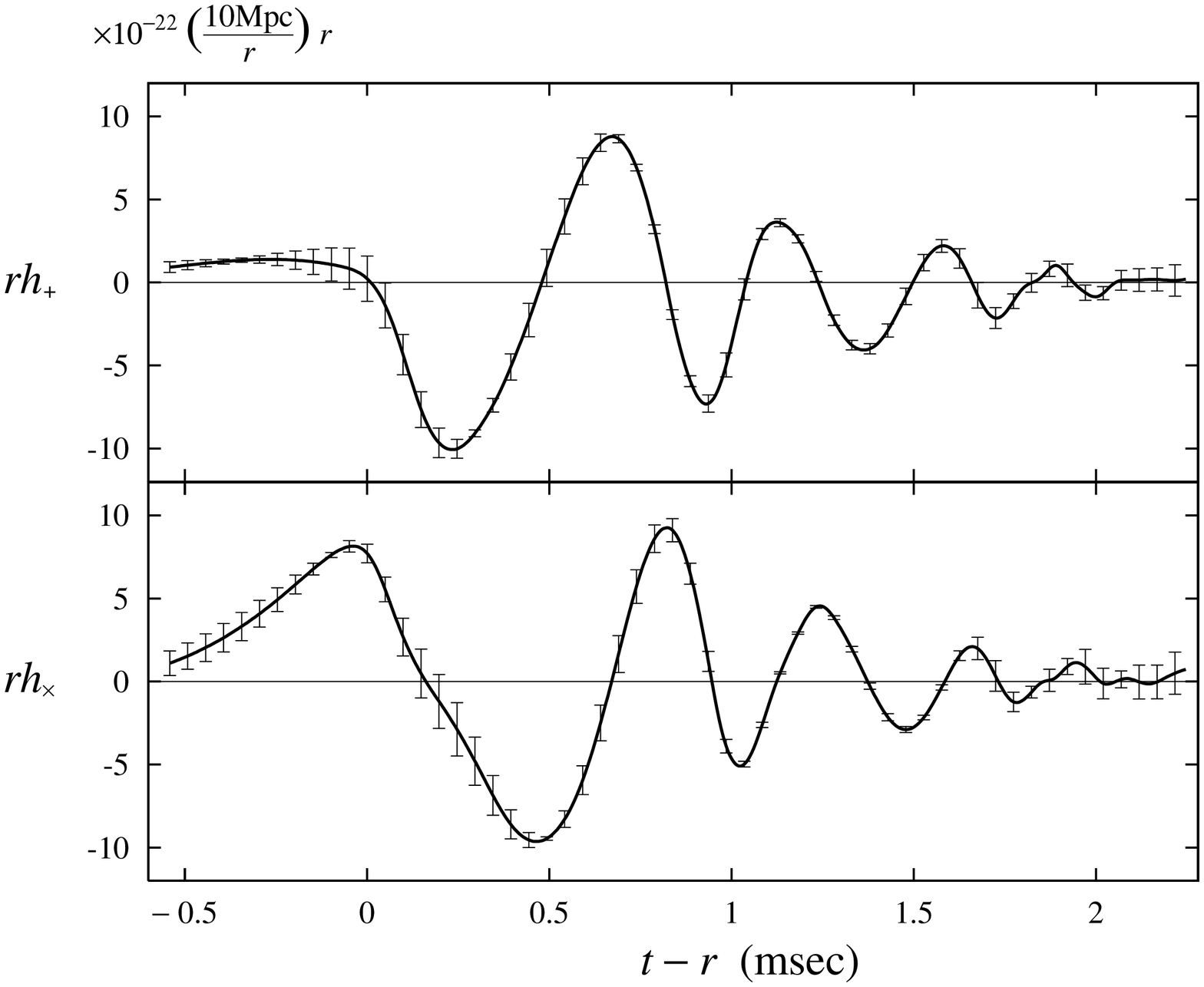}}
  \caption{Wave forms $r h_{+,\times}$ along $z$-axis as a function of
    $t-r$. The curves are averages of $r h_{+,\times}$ estimated at $r
    = 110 \sim 200 \sol$ and error bars denote $2\sigma$.}
    \label{fig:wave}
  \end{minipage}
  \hfill
  \begin{minipage}{.45\textwidth}
  \centerline{\includegraphics[width=\textwidth]{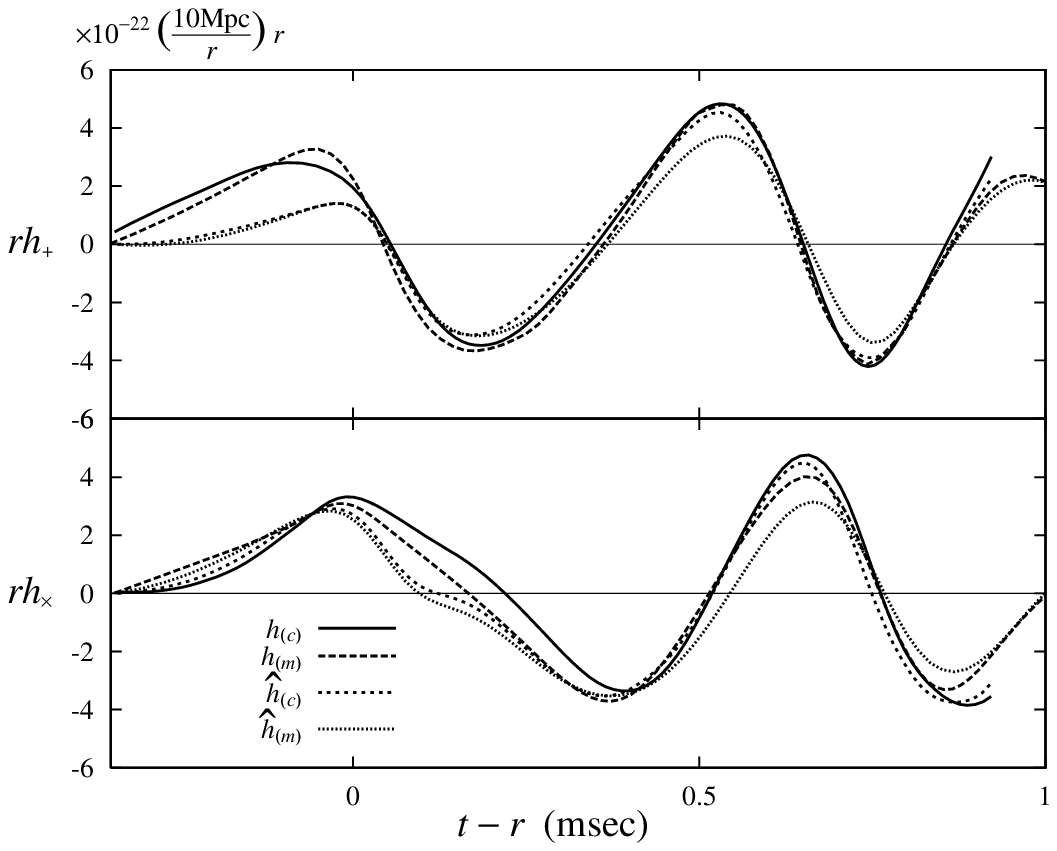}}
  \caption{The comparison $h_{+,\times}$ defined by
    Eq.(\ref{eq:defhpx}) and $\widehat{h}_{+,\times}$ 
    defined by Eq.(\ref{eq:hpx}) obtained with the conformal slicing
    (c) and the maximal slicing (m).}
    \label{fig:wavec}
  \end{minipage}
\end{figure}
The resultant wave form is shown in Fig.\ref{fig:wave}. The curves
represent the average of $h_+$ and $h_\times$ calculated at $r = 110,
120, \cdots 200\sol$ and twice the dispersion $2\sigma$ is shown as
error bars.

Here we define $\widehat{h}_+$ and $\widehat{h}_\times$ as
\begin{equation}
  \label{eq:hpx}
  \widehat{h}_+ = \frac{1}{2} \left( h_{xx} - h_{yy} \right)
  \quad
  \mbox{and} \quad
  \widehat{h}_\times  =  h_{xy},
\end{equation}
\begin{wrapfigure}{r}{\halftext}
  \centerline{\includegraphics[width=.4\textwidth]{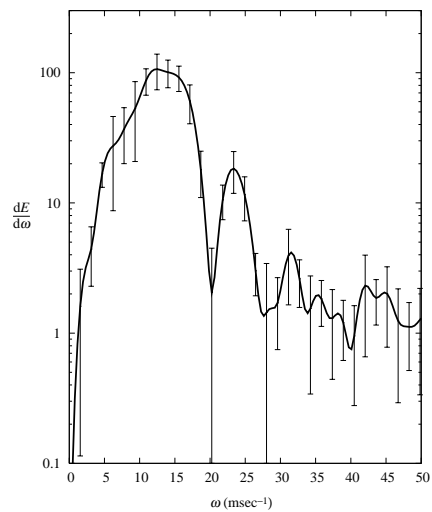}}
  \caption{The energy spectrum of the gravitational waves plotted in
    Fig~\ref{fig:wave}.
    The curves are averages of $\mbox{d}E/\mbox{d}\omega$ estimated at
    $r = 110 \sim 200 \sol$ and error bars denote $2\sigma$.}
    \label{fig:spec}
\end{wrapfigure}
respectively, where $h_{ij} = \phi^{-4} \gamma_{ij} - \delta_{ij}$ and
$\phi = \left( \mbox{det} (\gamma_{ij}) \right)^{\frac{1}{12}}$. The
pseudo-minimal distortion condition demanding $\partial_t (\partial_j
h_{ij}) = 0$ guarantees $\widehat{h}_{+,\times}$ to be
transverse-traceless if $\partial_j h_{ij} = 0$ at $t = 0$. It is our
case since we assumed the initial three-metric to be conformal flat,
$h_{ij} = 0$. Then they can be considered as the gravitational waves
on $z$-axis in the conformal slicing, while they include gauge
dependent modes in the maximal slicing.\cite{SN92} To compare
$\widehat{h}_{+,\times}$ with $h_{+,\times}$ for the conformal slicing
as well as the maximal slicing, we perform numerical simulation for a
coalescing binary of two $M=1.0\sol$ neutron stars. As shown in
Fig.~\ref{fig:wavec}, they almost coincide with each other, while a
small deviation is found in $\widehat{h}_{+,\times}$ in the maximal
slicing. Then we found that the gauge mode in $\widehat{h}_{+,\times}$
is small even in the maximal slicing.

Finally to investigate a possibility that the excitation of the
quasi-normal modes can be seen by the numerically calculated waves, we
evaluated the energy spectrum of the gravitational waves, which is
given by
\begin{equation}
  \frac{\mbox{d}E_{\rscript{GW}}}{\mbox{d}\omega}= \frac{1}{32\pi}
  \sum_{l,m} \omega^2 \left( \left| \Psi^{(e)}_{lm\omega}(r) \right|^2
  + \left| \Psi^{(o)}_{lm\omega}(r) \right|^2 \right),
\end{equation}
where $\Psi^{(I)}_{lm\omega}(r)$ is the Fourier transformation of
$\Psi^{(I)}_{lm}(t,r)$. Figure \ref{fig:spec} shows the energy
spectrum of the waves plotted in Fig.~\ref{fig:wave}. The fundamental
frequency of $l = 2$ for a Schwarzschild black hole of mass $2.8\sol$
is ${\omega}=25~\mbox{msec}^{-1}$. A peak near this frequency appears
in Fig.~\ref{fig:spec}. Unfortunately, however, rotating angular
frequency just when the merger of the stars finishes is
$12~{\sim}~15~\mbox{msec}^{-1}$ and thus they will radiate the waves
of frequency near ${\omega}=25~\mbox{msec}^{-1}$.  So that more
precise calculation is necessary to discuss whether this peak
corresponds to the emission of the quasi-normal mode of the formed
black hole.

Numerical computations were carried out on SR8000/F1 at Hight Energy
Accelerator Research Organization(KEK) and on VPP5000 at the
Astronomical Data Analysis Center of the National Astronomical
Observatory Japan(NAO).  This work was in part supported by
Grant-in-Aid for Scientific Research (C), No.13640271, from
Japan Society for the Promotion of Science, by the Supercomputer
Project No.097(FY2003) of KEK and by the Large Scale Simulation
Project yko12a of NAO.

%

\end{document}